\documentstyle[emulateapj,onecolfloat]{article}
\submitted{Submitted to ApJL.}

\newcommand{\mpc}{{\rm\,Mpc}}
\newcommand{\impc}{{\rm\,Mpc}^{-1}}
\newcommand{\hmpc}{h^{-1}{\rm\,Mpc}}
\newcommand{\ihmpc}{h{\rm\,Mpc}^{-1}}
\newcommand{\kmsmpc}{{\rm\ km\ s^{-1}\ Mpc}^{-1}}
\newcommand{\kmax}{k_{\rm max}}
\newcommand{\etal}{et al.}
\newcommand{\om}{\Omega_m}
\newcommand{\ob}{\Omega_B}
\newcommand{\ol}{\Omega_\Lambda}
\newcommand{\oK}{\Omega_K}

\newcommand{\tskip}{\tablevspace{3pt}}

\begin{document}
\twocolumn[%
\title{Cosmic Complementarity: $H_0$ and $\om$ from Combining\\
CMB Experiments and Redshift Surveys}
\author{Daniel J. Eisenstein, Wayne Hu, and Max Tegmark\altaffilmark{1}
}
\affil{Institute for Advanced Study, Princeton, NJ 08540}

\begin{abstract}
We show that upcoming CMB satellite experiments and large redshift surveys
can be used together to yield 5\% determinations of $H_0$ and $\om$,
an order of magnitude improvement over CMB data alone.
CMB anisotropies provide the sound horizon at recombination
as a standard ruler.  For
reasonable baryon fractions, this scale is imprinted on the galaxy
power spectrum as a series of spectral features. 
Measuring these features in redshift space determines 
the Hubble constant, which in turn yields $\om$ once
combined with CMB data.
Since the oscillations in both power spectra are frozen in at recombination,
this test is insensitive to low-redshift cosmology.
\end{abstract}

\medskip]

\altaffiltext{1}{Hubble Fellow}
\section{Introduction}
In the usual cosmological paradigm, 
the cosmic microwave background (CMB) contains a vast amount of
information about cosmological parameters (\cite{Hu97}\ 1997).  With
upcoming experiments, most notably the two satellite missions 
MAP\footnote{http://map.gsfc.nasa.gov}
and Planck\footnote{http://astro.estec.esa.nl/SA-general/Projects/Planck}, 
detailed measurements of the angular power spectra of
its anisotropy and polarization may accurately determine many cosmological
parameters (\cite{Jun96}\ 1996; \cite{Bon97}\ 1997; \cite{Zal97}\ 1997).
However, certain changes in the cosmological parameters can conspire
to leave the CMB power spectra unchanged, resulting in degenerate
directions in the parameter space (\cite{Bon94}\ 1994, 1997;
\cite{Zal97}\ 1997; \cite{Hue98}\ 1998).  For example, since 
the Hubble constant $H_0$
and the matter density $\om$ can be varied while keeping the
angular diameter distance and the matter-radiation ratio fixed, their
values remain uncertain but highly correlated.
Such degeneracies must be broken with cosmological information 
from other sources.

Upcoming redshift surveys for the study of large-scale structure
hold the potential for resolving this issue.  In particular, the 
2dF survey\footnote{http://meteor.anu.edu.au/$\sim$colless/2dF}
and the Sloan Digital Sky Survey 
(SDSS)\footnote{http://www.astro.princeton.edu/BBOOK}
should measure the galaxy power spectrum
on large enough scales to allow detailed comparisons to the
mass power spectra predicted by cosmological theories.  In this {\it Letter}
and a companion paper (Eisenstein, Hu, \& Tegmark 1998, hereafter \cite{EHT}),
we explore the potential of combining redshift surveys and CMB anisotropy
data for the purpose of parameter estimation.
Here, we
focus on the dramatic improvement possible in the measurement of 
$H_0$ and $\om$.  Neither data set yields tight limits by itself,
yet together they could yield errors better than 5\% on $H_0$ and
10\% on $\om$.

The key to this improvement is the presence of features in the matter
power spectrum on scales exceeding $60 h^{-1}\mpc$.  With a non-negligible 
baryon fraction, the acoustic oscillations
that exist before recombination are imprinted not only on CMB anisotropies
but also on the linear power spectrum
(\cite{Hol89} 1989, \cite{Hu96} 1996, \cite{Eis98a}\ 1998a).  
CMB anisotropies accurately calibrate
their characteristic length scale; measurement of this standard
ruler in the redshift survey power spectrum yields $H_0$.  With
this added information, the CMB returns a significantly
more precise measure of $\om$.

\section{Methodology}
\label{sec:Methodology}

We seek to quantify the potential sensitivity of these data
sets to various cosmological parameters.  For this, we use
the Fisher matrix formalism (see \cite{Teg97b}\ 1997 for a review), which  
yields a lower limit on the statistical errors on 
cosmological parameters achievable by a set of experiments.  
This formalism operates within the context of a
parameterized cosmological model.  For this we use a 12-variable 
parameterization of the adiabatic CDM model,
described in detail in \cite{EHT}.
It includes cold dark matter, baryons, massive neutrinos, a
cosmological constant $\ol$, curvature $\oK$ ($\equiv1-\ol-\om$), the 
Hubble constant $H_0\equiv100h\kmsmpc$, a reionization optical depth, and
a primordial helium fraction.   
It assumes an initial scalar
power spectrum $P_i(k)\propto k^{n_S+\alpha\log(k/k_p)}$
($k_p\equiv0.025\impc$) with tilt $n_S$, a logarithmic running of
the tilt $\alpha$, and an unknown amplitude as well as
scale-invariant tensor contributions with an unconstrained amplitude.  
Finally, it allows an
unknown linear bias to adjust the galaxy power spectrum relative to the
mass ($P_{\rm gal}=b^2P_{\rm mass}$).  
All of the above parameters are determined simultaneously from the data.

For CMB anisotropies, we use the experimental specifications of the
MAP and Planck satellites for temperature and polarization (\cite{EHT}).
We assume that foregrounds and systematics
can be eliminated with negligible loss of cosmological information.
For large-scale
structure, we use the projected specifications of the
Bright Red Galaxy (BRG) sample of the SDSS to determine
its sensitivity to the linear power spectrum (\cite{Teg97a} 1997). 
On small scales, the observed power spectrum reflects non-linear effects
and galaxy formation issues.
We therefore employ only wavenumbers less than 
$\kmax=0.1\ihmpc$ under the assumption that 
the linear power spectrum on these scales can be reconstructed (up to 
the unknown linear bias) from the observed quasi-linear information
with no adjustment to the error bars.  
We vary $\kmax$ in \S\ \ref{sec:Caveats}. 

\section{CMB Results}
\label{sec:CMBResults}

Parameter degeneracies occur when 
changes in the model parameters leave the 
power spectra essentially unchanged relative to the size of the
experimental uncertainties.  In particular, since cosmic
variance is substantial at large angular scales, 
changes that affect large angles while leaving the acoustic peaks unchanged 
will be difficult to detect.

The angular diameter distance $d_A$
to the last scattering surface contains the most important degeneracy 
for the present discussion.
The CMB acoustic peaks are a high-redshift pattern  viewed at distance $d_A$.
The pattern may be held 
fixed by keeping $\om h^2$ and the baryon density $\ob h^2$ 
constant.  However, $d_A$ depends 
on the low-redshift effects of a cosmological constant or
curvature.
Changing $\ol$ and $\oK$ so as to keep the 
angular diameter distance constant leaves the acoustic peaks unchanged.
Only large-angle ($\ell\lesssim 50$) gravitational redshift effects or 
small-angle ($\ell\gtrsim1000$) gravitational lensing effects 
can resolve this ambiguity.

In short, the CMB data sets will yield precision information on
the physical properties at high redshift, notably $\om h^2$,
$\ob h^2$ and $d_A(\om h^2,\Omega_\Lambda,\Omega_K)$, but not
on $H_0$ and $\om$ individually.  
A similar situation occurs in quintessence models with trade-offs between 
$\Omega_Q$ and the equation of state of the $Q$-field (\cite{Hue98} 1998).

In Table \ref{tab1}, we present the error bars on $H_0$ and $\om$
attainable by upcoming CMB satellite experiments within our
12-dimensional parameter space.  One sees that when varying both
$\ol$ and $\oK$, the constraints on $H_0$ and $\om$ are poor,
although polarization information does provide considerable help.
Even if one assumes a flat cosmological model ($\oK=0$),
MAP with its partial coverage of the acoustic peaks 
will not yield tiny errors on $H_0$ and $\om$. 

\begin{figure}[tb]
\centerline{\epsfxsize=3.5in\epsffile{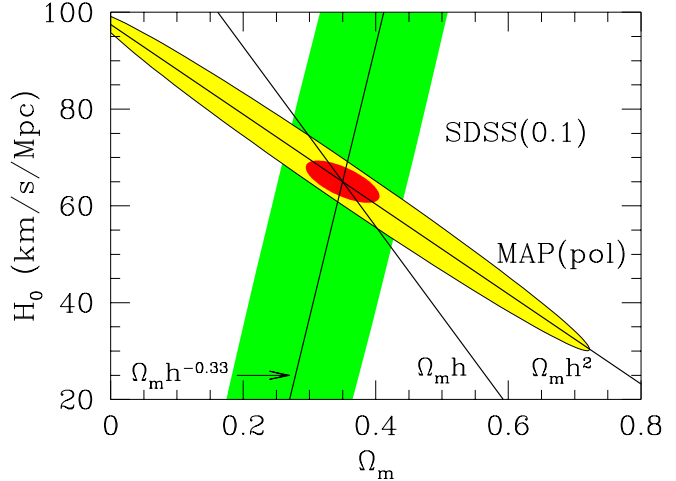}}
\figcaption{The 68\% allowed regions for MAP (with polarization) alone,
SDSS ($\kmax=0.1\ihmpc$) alone, and the two combined.  Lines in the
direction of constant $\om h^2$, $\om h$, and $\om h^{-0.33}$ are shown.
\label{fig1}}
\end{figure}

There are several caveats.  
The Fisher matrix expansion of the likelihood
function is not accurate for large steps in parameter space, meaning
that large error bars accurately detect a degenerate
direction but may inaccurately reflect its magnitude (\cite{Zal97}\ 1997). 
One artifact of this is that the ellipses in Figure \ref{fig1} follow
straight lines rather than curves, e.g. constant $\om h^2$ in the CMB case.
Moreover, the limits are slightly overestimated
in the case of Planck because we have not included gravitational
lensing (\cite{Sel96a}\ 1996; \cite{Met97}\ 1997),
by which the differences in growth factor between otherwise degenerate
models will alter the small angle power spectra. 
Nevertheless, the point remains that the CMB alone will not constrain
$H_0$ and $\om$ to a level capable of strong consistency 
checks against other cosmological tests.

\begin{table}[tb] \footnotesize
\caption{\label{tab1}}
\begin{center}
{\sc Errors on $H_0$ and $\om$ for $\Lambda$CDM\smallskip}
\begin{tabular}{lcccc}
\tableline\tableline\tskip
&\multicolumn{2}{c}{$\Delta H_0$}&
	\multicolumn{2}{c}{$\Delta \om$}\\
Experiment&General&Flat&General&Flat\\
\tskip\tableline\tskip
MAP (no Pol.) & 135 & 15 & 1.4 & 0.23 \\ 
with SDSS & 3.0 & 2.5 & 0.042 & 0.037 \\[5pt]
MAP (with Pol.) & 23 & 6.7 & 0.25 & 0.10 \\ 
with SDSS & 2.9 & 2.4 & 0.037 & 0.036 \\[5pt]
Planck (no Pol.) & 113 & 5.3 & 1.2 & 0.079 \\ 
with SDSS & 2.5 & 2.3 & 0.035 & 0.035 \\[5pt]
Planck (with Pol.) & 13 & 1.6 & 0.14 & 0.024 \\ 
with SDSS & 2.2 & 1.4 & 0.027 & 0.020 \\
\tskip\tableline
\end{tabular}
\end{center}
{NOTES.---%
The fiducial model has $\om=0.35$, $H_0=65$ km s$^{-1}$ Mpc$^{-1}$, 
and $\ob=0.05$.  We use $\kmax=0.1\ihmpc$ for SDSS.
Errors are 1-$\sigma$; $\Delta H_0$ errors are in units of $\kmsmpc$.
General: $\oK$ estimated from data.  Flat: $\oK=0$ by fiat.}
\end{table}

\section{Results with Redshift Surveys}
\label{sec:BRGResults}

\subsection{Linear Analysis}

When we include the Fisher information matrix from SDSS, the error
bars on $H_0$ and $\om$ drop by an order of magnitude.
In Table \ref{tab1}, we see that for a fiducial $\Lambda$CDM model, the errors on 
$H_0$ are below $3\kmsmpc$ while those on $\om$ are around 0.035.
\cite{EHT} discuss improvements on other parameters of the model; however,
none are nearly as dramatic.  Figure \ref{fig1} displays the situation.
Note that the results are roughly independent of whether polarization
information is available or not.

The reason for this dramatic improvement lies with the
baryons.  Table \ref{tab2} shows the error bars on $H_0$ for 
a sequence of fiducial models with increasing baryon fraction; 
those on $\om$ behave similarly.  Increasing the baryon fraction 
from $\sim\!1$\% to $\sim\!15$\% results in a dramatic
increase in the information provided by SDSS. 
A baryon fraction exceeding $10\%$ is
strongly indicated by cluster gas fractions 
(\cite{Whi93}\ 1993; \cite{Dav95}\ 1995; 
\cite{Whi95}\ 1995; \cite{Evr97}\ 1997).

\begin{figure}[tb]
\centerline{\epsfxsize=3.5in\epsffile{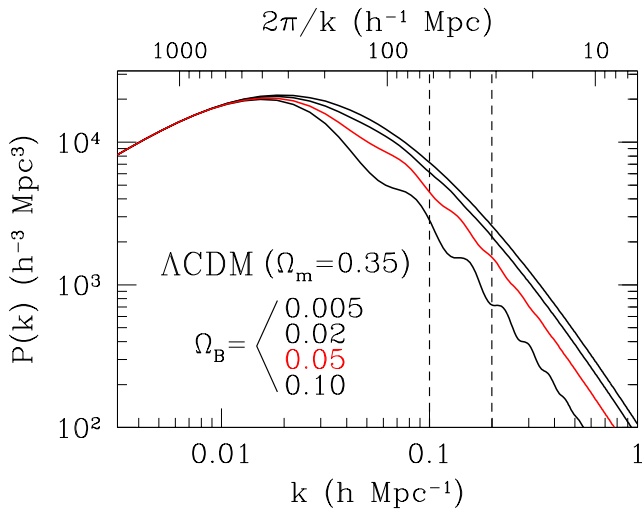}}
\figcaption{Power spectra for 4 $\Lambda$CDM models, showing a 
progression of $\ob$ (0.005, 0.02, 0.05, 0.1).  The spectra
are normalized on large scales.  
\label{fig2}}
\end{figure}

\begin{table}[tb]\footnotesize
\caption{\label{tab2}}
\begin{center}
{\sc Errors on $H_0$ as Function of $\ob$\smallskip}
\begin{tabular}{cccc}
\tableline \tableline\tskip
$\ob$ & $\ob/\om$ & MAP (w/ Pol) & with SDSS \\
\tskip\tableline\tskip
0.005 & 1.4\% & 36 & 12 \\
0.02 & 5.7\% & 29 & 9.2 \\
0.05 & 14\% & 23 & 2.9 \\
0.10 & 29\% & 24 & 1.3 \\
\tskip\tableline
\end{tabular}
\end{center}
{NOTES.---%
Same parameters as Tab.~\protect\ref{tab1} save for the baryon fraction.}
\end{table}

As the baryon fraction increases, significant acoustic
oscillations develop in the matter power spectrum 
(see Fig. \ref{fig2} and \cite{Hu96} 1996, \cite{Eis98a}\ 1998a).  
There is a characteristic scale of these oscillations known as
the sound horizon;
morphologically, the power spectrum acquires a sharp break near
the sound horizon and a series of oscillations marking harmonics
of this scale.  The size of the sound horizon can be calculated
given knowledge of the physical conditions at high redshift,
particularly $\om h^2$ and $\ob h^2$.  Since these are exactly
the quantities that are well-constrained by the relative heights
of the CMB acoustic peaks,
we can accurately infer the scale of
these features in real space.  Its measurement in the redshift-space 
power spectrum then yields an accurate measure of $H_0$.

At low baryon fractions, the SDSS power spectrum still reduces the
error bars on $H_0$ and $\om$.  This results from the
one scale left in the matter power spectrum, that of the horizon
at matter-radiation equality.  In redshift space, this yields a 
measure of $\Gamma\equiv\om h$, with considerable inaccuracies 
due to confusion with scalar tilt.  However, the fact that the 
SDSS ellipse in Figure \ref{fig1} lies along a line very different than
constant $\om h$ indicates that at moderate baryon fractions
this feature is not providing the primary leverage on $H_0$.  Note
that the break-and-oscillation morphology of the baryonic features 
cannot be mimicked by the effects of tilt or massive neutrinos.

In short, baryonic features yield a standard ruler, whose length
can be accurately inferred by the CMB and be measured in
redshift space using large redshift surveys.  The comparison of
lengths yields $H_0$; combining this with $\om h^2$ gives $\om$.
This inference is independent of the angular diameter distance
to last scattering (provided that the peaks are visible at all!),
so this method will function regardless of cosmological constant,
spatial curvature, or more exotic smooth components
(e.g., \cite{Tur97}\ 1997).  With $\om$ known, the location of the CMB peaks
and information from supernovae Ia (\cite{Per98}\ 1998; \cite{Rie98}\ 1998)
may be focused on distinguishing these low-redshift effects
(\cite{Teg98b}\ 1998b; \cite{Hu98b} 1998b).

\subsection{Non-Linearities}
\label{sec:Caveats}

Baryonic oscillations are a feature of the {\it linear} power spectrum.
Non-linear evolution erases these signatures, even
at second order in perturbation theory (e.g., \cite{Jai94}\ 1994).  
Hence we should test the extent to which the above results depend upon our use
of linear theory. 

Heretofore, we have assumed that we could use the linear power spectrum
on scales longward of $\kmax=0.1\ihmpc$.  
This is close to the point at which non-linearities will smooth the
oscillations.  We therefore display in Table \ref{tab3} the effect of
altering $\kmax$, again simply ignoring information on all smaller scales.  
For several fiducial
models, we find that moving $\kmax$ from $0.1\ihmpc$ to $0.2\ihmpc$
decreases the errors on $H_0$ and $\om$ by about a factor of 2.5.
However, these gains saturate as one extends $\kmax$ to $0.4\ihmpc$; 
the acoustic oscillations there are
of such small amplitude that little information is gained.

We also consider an alternative formulation in which we model the 
matter power spectrum by a fitting formula (\cite{Eis98b}\ 1998b) that
includes the break at the sound horizon but not the
oscillations.  
The resulting errors are shown in Table \ref{tab3}.  
For $\kmax\gtrsim0.08\ihmpc$, the performance is significantly
worse than that achieved with the actual linear power spectrum.
This is close to the location of the first bump in this fiducial
model.  Note that including the featureless power spectrum on scales
from $0.1\ihmpc$ to $0.4\ihmpc$ adds very little additional information
on $H_0$ or $\om$.  

\begin{table}[bt]\footnotesize
\caption{\label{tab3}}
\begin{center}
{\sc Errors on $H_0$ for Differing SDSS Assumptions\smallskip}
\begin{tabular}{ccccc}
\tableline \tableline\tskip
& \multicolumn{2}{c}{$\Delta H_0$} & 
	\multicolumn{2}{c}{$\Delta \om$} \\
$\kmax$ & $P(k)$ & $P_S(k)$ & $P(k)$ & $P_S(k)$ \\
\tskip\tableline\tskip
MAP alone & 23 & 23 & 0.25 & 0.25 \\
0.025 & 16 & 15 & 0.17 & 0.16 \\
0.05 & 9.7 & 10.7 & 0.098 & 0.11 \\
0.1 & 2.9 & 10.0 & 0.037 & 0.11 \\
0.2 & 1.2 & 9.0 & 0.016 & 0.10 \\
0.4 & 0.9 & 8.6 & 0.014 & 0.10 \\
\tskip\tableline
\end{tabular}
\end{center}
{NOTES.---%
MAP with polarization has been taken in each case.
Limits with the actual linear power spectrum $P(k \le \kmax)$ 
are compared with those from
a smoother analytic form $P_S(k \le \kmax)$ (see text). Same model and notation
as Tab.~\protect\ref{tab1}.}
\end{table}

We therefore conclude that detection of at least the first of the
acoustic oscillations ($k\approx0.07\ihmpc$ in this model) is 
critical to enabling a precision measure of $H_0$ and $\om$.
Detecting additional peaks improves the possible error bars but with
diminishing returns because the oscillations damp down in amplitude.
Cosmological simulations normalized to
the cluster abundance suggests that the first peak will indeed be unaffected
but that higher peaks will be smeared out (\cite{Mei98}\ 1998). 

\section{Discussion}
\label{sec:Discussion}

Detection of acoustic oscillations in the matter power spectrum would
be a triumph for cosmology, as it would confirm the 
standard thermal history and the gravitational instability paradigm.
Moreover, because the matter power spectrum displays these 
oscillations in a different manner than does the CMB,
we would gain new leverage on cosmological parameters.
In particular, we have shown in this {\it Letter} that 
the combination of power spectrum measurements from a galaxy
redshift survey with anisotropy measurements from CMB satellite experiments 
could yield a precision measurement of $H_0$ and $\om$.

The potential measurement of $H_0$ and $\om$ depends critically 
on the ability of the redshift survey to detect
the baryonic features in the linear power spectrum.  The best possible
error bars are a strong function of the baryon fraction but are
surprisingly good even if the fraction is $\sim\!10$\%, roughly the minimum
implied by cluster observations.  For such cases, the fractional limits
achievable with the SDSS are 5\% for $H_0$ and 10\% for $\om$ if
only the first acoustic peak in $P(k)$ is detected.  Detecting the
smaller-scale peaks could allow an additional factor of 3 refinement;
the exact limits would depend upon the scale at which non-linear
effects smooth out the power spectrum.  The results depend only mildly
on the details of the CMB experiment:  we find only slight gains
as our presumed CMB data set improves from MAP without polarization to
Planck with polarization.  While we have quoted numbers for SDSS,
it is possible that the 2dF survey will be able to make significant
progress on the detection of features in the power spectrum on very
large scales.
Unfortunately, the hints of excess power on $100\hmpc$ scales are
not likely to be due to baryons (\cite{Eis98c} 1998).

We have treated the galaxy power spectrum assuming linear bias on
large scales.  There is some theoretical motivation for this
(\cite{Sch98}\ 1998); moreover, if bias tends towards unity
as structure grows (\cite{Fry96} 1996; \cite{Teg98a}\ 1998), 
then scale dependences
in the bias at the time of formation will be suppressed.
Most importantly, this method of measuring $H_0$ and $\om$
depends upon extracting an oscillatory feature from the power spectrum. 
While one cannot prove that scale-dependent bias should be monotonic
on the largest scales, this seems more likely than an oscillation!
Finally, the assumption of linearity can be tested by constructing
the power spectrum with different types of galaxies 
(e.g., \cite{Pea97}\ 1997); 
future redshift surveys will allow this to be done on very large
scales with good statistics.

The method proposed here yields $H_0$ independent of local distance
measurements and $\om$ without the complications inherent in
dynamical methods.  In that, it is free of many confusing astrophysical
problems.  On the other hand, it does depend upon restricting oneself
to a class of models with observable acoustic oscillations in 
both CMB anisotropies and the galaxy power spectrum.
This assumption will be definitively tested from the data itself.

If the method described in
this {\it Letter} can yield tight constraints on $H_0$ and $\om$,
it will then be very important to compare these to other measurements
of these quantities.  In the coming decade, there will be a number
of paths toward a precision measure of $H_0$, such as the local
distance ladder (e.g., \cite{Fre98}\ 1998), gravitational lensing 
(e.g., \cite{Bla96}\ 1996), and
the S-Z effect (e.g., \cite{Coo98}\ 1998).  
Similarly, good estimates of $\om$ may be 
possible from velocity fields (e.g., \cite{Dek97}\ 1997), cluster evolution 
(\cite{Car97a}\ 1997a; \cite{Bah97}\ 1997), and $M/L$ measurements
(e.g., \cite{Car97b}\ 1997b).  
If the results
from these diverse sets of measurements are found to agree, 
we will have a secure foundation upon which to base our cosmology.

Acknowledgements: Numerical power spectra were generated with CMBFAST
(\cite{Sel96b} 1996).
We thank Martin White for useful discussions.
D.J.E.\ is supported by a Frank and Peggy Taplin Membership;
D.J.E.\ and W.H.\ by NSF-9513835;  
W.H.\ by the Keck Foundation and a Sloan Fellowship;
M.T.\ by NASA through grant NAG5-6034 and Hubble
Fellowship HF-01084.01-96A from STScI, operated by AURA, Inc.
under NASA contract NAS4-26555.

\end{document}